# Mechanisms of near-surface structural evolution in nanocrystalline materials during sliding contact

Zhiliang Pan[a], Timothy J. Rupert[ab*]

[a] Department of Mechanical and Aerospace Engineering, University of California, Irvine, California 92697, USA

[b] Department of Chemical Engineering and Materials Science, University of California, Irvine, California 92697, USA

*To whom correspondence should be addressed: trupert@uci.edu

## ABSTRACT

The wear-driven structural evolution of nanocrystalline Cu was simulated with molecular dynamics under constant normal loads, followed by a quantitative analysis. While the microstructure far away from the sliding contact remains unchanged, grain growth accompanied by partial dislocations and twin formation was observed near the contact surface, with more rapid coarsening promoted by higher applied normal loads. The structural evolution continues with increasing number of sliding cycles and eventually saturates to a stable distinct layer of coarsened grains, separated from the finer matrix by a steep gradient in grain size. The coarsening process is balanced by the rate of material removal when the normal load is high enough. The observed structural evolution leads to an increase in hardness and decrease in friction coefficient, which also saturate after a number of sliding cycles. This work provides important mechanistic understanding of nanocrystalline wear, enabled by the quantitative description of grain structure evolution.

# I. INTRODUCTION

Nanocrystalline materials are excellent wear-resistant coatings due to their high strength and hardness [1]. Extremely small grain sizes provide limited space for traditional dislocation activities such as tangling, forest interactions, and pile-up formation, with plastic deformation now dominated by new mechanisms that emphasize the grain boundaries either as sources and sinks for dislocations or even as direct carriers through grain boundary sliding and rotation [2-6]. As a result, nanocrystalline materials often experience unique wear-induced structural evolution near the contact surface. For coarse-grained materials, grain sizes near the contact surface are typically refined by the dislocation cell structures formed in the local plastic deformation during the sliding process [7]. However, nanocrystalline materials have been observed to experience grain coarsening during the sliding wear process [8,9]. This evolution is especially interesting because it has been linked to unexpectedly high wear resistance. Wear rate, or the volume of worn material per unit sliding distance, usually follows the Archard model [10] and is inversely proportional to the hardness of the material prior to testing, meaning harder materials wear less. Rupert and Schuh [9] found that the Archard model no longer captures the wear behavior of electrodeposited nanocrystalline Ni-W, with these materials having wear resistance that is higher than that predicted by their as-deposited hardness due to the evolution of the surface microstructure and properties.

The type and extent of wear-driven structural evolution of nanocrystalline materials observed in experiments can vary with the material's starting grain size, the choice of coating material, sliding speed, contact pressure, and other factors. For Ni-W, when the grain size is extremely small (less than 15 nm), a grain growth layer forms on top of the ultrafine nanocrystalline matrix, with or without a transition region comprised of a grain size gradient in between depending on the contact load and therefore the applied stress [8,9,11]. At larger grain sizes, no substantial structural change is observed during wear in Ni-W [9] or Mg [12]. However, grain growth has been observed in Ni-Fe when the initial grain size is 34 nm, forming a layer of large grain sizes sandwiched between an ultrafine nanocrystalline thin layer right underneath the contact surface and the nanocrystalline matrix below [13]. Another multilayer configuration was found after the sliding wear of nanocrystalline Ni with grains in the range of 20–100 nm [14], except that the boundary between the ultrafine nanocrystalline layer and the large grain layer was not flat, but rather wavy due to the formation of a vortex-like structure. Another multilayer is found in a mixed nanostructured Cu that contains both nanoscale twin lamellae and typical nano-sized grains, with a vortex structure right underneath the contact surface, followed by a ultrafine grained layer, and then finally the nanocrystalline matrix [15].

The discussion above shows that structural evolution near contact surfaces is widespread in nanostructured materials, yet this behavior can only be fully understood if the detailed mechanisms behind the evolution are uncovered. Experiments struggle to accomplish this goal, as measurements of surface structure are often made only before and after the wear process. In addition, it is particularly difficult to measure spatial gradients in grain size when both grains and the distance over which their size changes is only in the nanometer range. Comparatively, atomistic simulations offer a much higher spatial and temporal resolution, since they are able to retrieve all of the atomic details during wear. Such techniques have been used to show that dislocation mechanisms, grain boundary activities, and deformation twinning control the plasticity of nanocrystalline Cu during a wear process [16], with the relative contribution depending on the scratching rate [17]. The formation mechanism of the displaced or removed material by the indenter during sliding process of nanocrystalline Cu has also been studied [18], showing that dislocation activities of some well-aligned slip systems are responsible for the bulging and fold formation. Li and Szlufarska [19] studied the dependence of wear on grain size in nanocrystalline Cu, finding that the grain size effect is coupled to the indenter tip size and identifying an optimum grain size that has the best wear resistance. The wear behavior of nanocrystalline SiC has also been studied with molecular dynamics, showing that grain boundary sliding is the primary deformation mechanism [20]. An alternative geometry was a focus of the work by Romero et al. [21], who simulated two nanostructured slabs that were pushed together and moved relative to one another. These authors found evidence of coarsening in a localized shear band, similar to earlier observations from molecular dynamics of nanocrystalline nanowire samples in tension [22]. Although this collection of prior work shows that atomistic simulations are a valuable tool for modeling wear, the evolution of surface structure and resulting mechanical properties, as well as the mechanisms responsible for this evolution, have not yet been quantitatively investigated in detail, especially as a function of time or sliding cycle. Reports of wear-induced coarsening have been largely qualitative to date, without a full description of grain size distribution and resultant texture (preferred crystal orientation).

In this work, we isolate and quantify structural evolution under constant-load sliding contact using nanocrystalline Cu as a model system. Grain growth was observed during the sliding process, with faster growth and a more discrete coarsened layer promoted by the application of higher normal loads. Quantitative analysis of the structural evolution is performed using a recently developed Grain Tracking Algorithm (GTA) [23-25], indicating that the structural evolution continues with increasing number of wear cycles by first forming a gradient structure and then saturating to a distinct coarsened layer that is balanced by the rate of material removal at high normal load. This work allows us to measure sub-surface damage in a quantitative way that differs from prior modeling studies, with results comparable with experimental observations.

## II. SIMULATION METHODS

Atomistic simulations were performed using the Large-scale Atomic/Molecular Massively Parallel Simulator (LAMMPS) code [26], with integration time step of 1 fs. An embedded-atom method potential was used to describe the interatomic interactions of Cu atoms [27]. Structural analysis and visualization of atomic configurations were performed using the open-source visualization tool OVITO [28], with local structural of each atom identified based on adaptive common neighbor analysis (CNA) [29,30]. For figures using CNA to show atom structure, face-centered cubic atoms are colored green, hexagonal close packed atoms are colored red, body-centered cubic atoms are colored blue, icosahedral atoms are colored yellow, and other atoms (usually grain boundaries or surfaces) are colored white. The starting configuration is a nanocrystalline sample created through Voronoi tessellation construction, with an average grain size of 5 nm, as shown in Figure 1. This fine grain size was chosen to be comparable to experimental reports in nanocrystalline Ni-W, where the most obvious structural evolution was found for alloys with starting average grain sizes of 3-5 nm [8,9]. The simulation cell is quasi two-dimensional and 45 nm long (X direction), 50 nm tall (Y direction), and 15 nm thick (Z direction), with the thickness chosen to make sure at least three grains are distributed in this direction. As a whole, the sample contains ~500 grains and ~3,000,000 atoms. The sample was first equilibrated with a conjugate gradient minimization technique to the minimum potential energy state, with periodic boundary conditions applied in all three directions. A Nose-Hoover thermo/barostat was then used to anneal the sample for 50 ps under zero pressure at 600 K, to relax any unrealistic grain shapes or grain boundaries associated with the Voronoi construction technique. After annealing, the sample was cooled down to 300 K over 50 ps and then kept at this temperature for another 20 ps. Thereafter, the periodic boundary condition was removed in the Y direction to create free surfaces and a 1 nm thick layer of atoms at the bottom was fixed. The sample was then relaxed for additional 20 ps at 300 K in a canonical ensemble to construct the final sample with the contact surface on top. The final average grain size increases to 5.5 nm due to slight grain growth during the annealing during sample preparation.

A cylindrical indenter was used to first indent at the center of the top surface and then scratch the sample in the X direction at different normal loads, as shown schematically in Figure 1, with the cylinder axis parallel to the sample thickness direction and the axis length equal to the sample thickness. The cylindrical indenter and quasi two-dimensional sample setup were chosen so that the contact stresses only vary as a function of one dimension: the depth into the sample in the Y direction. This is useful when trying to quantify microstructural evolution, since grain size and other features can be calculated as a function of depth. The indenter radius is 8 nm so that it can span at least 3 grains in the X direction. The interaction between atoms and the indenter is modeled with a repulsive potential [31], with the exerted repulsive force on the indenter calculated by $F(r) = -K(r-R)^2$ when $r \leq R$ and $F(r) = 0$ when $r > R$. Here $r$ is the distance from the atom to the center axis of the indenter and $R$ the indenter radius. $K$ is the force constant, specified to be 1602 GPa (10 eV/Å$^3$) in this work. This repulsive force model and $K$ equal or close to this value have been widely used and validated in atomistic simulations of indentation in metallic systems [31-33].

A load-controlled indentation and sliding methodology was created to allow for multiple sliding passes. The vast majority of atomistic simulations of sliding wear in the literature use a displacement-controlled technique, where an indentation depth is defined ahead of time and kept constant. This is problematic when trying to observe cyclic evolution, as the indenter tip will gradually lose contact with the surface as more material is removed. In contrast, Hu et al. [34]

were able to simulate a constant load by applying load directly on the indenter top, where the indenter is modeled with a cluster of atoms and atoms at the top are held as a rigid body. Other similar techniques have been reported in Refs [35,36]. To maintain a constant contact load in this study, the vertical position of the indenter was adjusted every time step during the indentation and sliding process, with atomic positions and velocities updated in a canonical ensemble at 300 K. The indenter was moved up if the load is higher than the target load and moved down otherwise. A maximum speed of 240 m/s was set for the correction to reduce the overshoot of the load. If overshoot does happen to occur, the maximum speed was reduced to 60 m/s to avoid large vibrations around the target load. The whole indentation process continues for 40 ps. Figure 2(a) shows how the load changes with indentation depth with a target load of 481 nN, corresponding to 300 eV/Å$^2$ in units more directly related to the repulsive potential. The load first increases slowly when the indenter gradually touches the materials, then increases rapidly when the indenter pushes into the material until the load reaches the target value and is held at this value. For the sliding portion of the simulation, the indenter was moved in the positive X direction at a speed of 100 m/s while maintaining the constant load. Figure 2(b) shows how the normal load changes during this sliding process. The load fluctuates around the target value, with the deviations being less than ±1% of the target load.

The sliding process continues up to 8 cycles. After each cycle, the grain size was quantitatively analyzed using the GTA code [24]. This code can identify grains based on the local crystallographic orientation of each atom as well as characterize grain boundaries by their misorientation and grain boundary plane normal [25], while also tracking such features throughout a simulation. Atoms are first categorized into grain interior (face-centered cubic) or defects atom based on the centrosymmetry parameter [31]. Then the crystallographic orientation of each crystalline atom is calculated based on its local environment. After that, grains are identified as the cluster of neighboring crystalline atoms between which the misorientation angle are less than 1°. Before GTA analysis, conjugate gradient energy minimization was used to remove thermal noise in atomic misorientation so that all atoms belonging to a grain are counted when using such a small angle cutoff. This is done to prevent the artificial identification of neighboring grains as one [24]. The size of each grain is then calculated as the diameter of a sphere that has the same volume with the grain. In this study, stacking faults in the grain interiors were counted as crystalline atoms, to avoid incorrectly identifying one grain as two in the case of a stacking fault that bisects a grain.

## III. RESULTS AND DISCUSSION

Figure 3 presents snapshots right before and after indentation with a normal load of 481 nN applied on the indenter and atoms colored according to CNA. Before indentation, the whole sample shows a typical polycrystalline configuration decorated by some twin boundaries (single layer of red atoms) and stacking faults (double or multiple layers of red atoms). After indentation, new stacking faults, as indicated by the black arrows, have appeared beneath the indenter due to the indentation process. Right after indentation, the contact pressure (defined as the applied normal load divided by projected contact area) between the indenter and the sample is 4.24 GPa. The indenter starts to move in the positive X direction from this indented configuration, which therefore corresponds to a sliding displacement of 0 nm. Figure 4 presents snapshots at higher sliding displacements during the first sliding cycle. At a sliding displacement of 10 nm, additional stacking faults appear both in front of and behind the indenter tip. The number of stacking faults continues to increase at higher sliding displacement, indicating an increased activity of Shockley partial dislocations. New twin boundaries, indicated by the black arrows, also begin to form at a sliding displacement of 20 nm, with the population increases with the increasing displacement. A small pileup of material in front of the indenter is observed in all frames. The first sliding cycle is completed when the indenter moves back to the center position and a new cycle starts as the indenter keeps moving in the positive X direction.

Figure 5 shows the distribution of stacking fault and twin boundary atoms along the Y position before and after indentation, as well as for different sliding displacements during the first wear cycle. This data supports the aforementioned qualitative description of the trends for these two defect structures. Stacking faults are identified as two or more planes of hexagonal close packed atoms and twin boundaries as an isolated plane of hexagonal close packed atoms. Right after indentation, the number of stacking fault atoms has increased significantly near the sample top when compared with the starting configuration. During the sliding process, the number of stacking fault atoms also increases with increasing sliding displacement, but the rate of change slows down when the first cycle is about to be completed. The change of twin boundary atoms follows the same trend as the stacking fault atoms, with the change occurring closer to the sample surface in this case, consistent with the observation that new twin boundaries appear after new stacking faults during the first sliding cycle.

Figure 6 presents snapshots after 1–6 sliding cycles. Both partial dislocations and deformation twins continue to be formed during the repetitive sliding process. At the end of the 5$^{th}$ cycle, a fivefold twin forms, as shown in the inset to Figure 6. This is not an isolated event, as another fivefold twin was formed at a distance 12 nm in the Z direction into the page (not visible in Figure 6). This type of twin can increase the yield strength of nanowires [37,38] and are unlikely to form during uniaxial deformation of nanocrystalline materials, but can be observed after severe plastic deformation [39]. Zhu et al. [40] have attributed such observations to the complex, three-dimensional stress state associated with severe plastic deformation, which is also found around the sliding indenter during the scratching process. The pileup asperity formed in front of the indenter is largely gone after the second cycle and the number of white atoms (most commonly grain boundary atoms) near the contact surface gradually decreases with increasing number of cycles, indicating that grain growth happens during the scratching process. The coarsened region also goes deeper into the material with more sliding cycles. It should be noted that both the grain size, grain shape, boundary network and preexisting stacking faults or twin boundaries at the lower part of the sample do not change in any meaningful way, indicating that structural evolution only occurs close to the contact surface.

Similar structural evolution and deformation features are also observed when the load is increased to 513 nN (contact pressure of 4.31 GPa right before sliding). However, when the load is increased to 529 nN (contact pressure of 4.25 GPa right before sliding), as shown in Figure 7, there is significantly more material removal and the pileup in front of the indenter grows with increasing cycle number. At the same time, the indenter goes deeper and deeper into the sample, indicating that material is being removed during this scratching process. During this removal process, the layer underneath the contact still shows evidence of partial dislocations, deformation twinning, and grain growth. Further increases to the applied normal load give more pileup and higher material removal rates. It is worth noting that the contact pressure at 529 nN is lower than that at 513 nN. This happens because there is more plastic deformation at 529 nN, which pushes the indenter so much deeper into the sample that the resulting increase in projected contact area outpaces the increase in applied normal load. This pressure drop could perhaps be used as an indicator for the transition from pure grain growth to material removal that is accompanied by grain growth. If the normal load is increased from a very low level, a similar pressure drop can be expected when the sample experiences a transition from pure elastic deformation to plastic deformation during the indentation process.

While we do not present the results in detail, scratching simulations were also performed on nanocrystalline Ni, with the interatomic interactions of Ni atoms described using another embedded-atom method potential [41]. An identical sample size and grain size was chosen, with higher normal loads since Ni is generally stiffer and stronger than Cu. Partial dislocation nucleation, deformation twinning, and grain growth that increased with increasing numbers of sliding cycles were also observed, indicating that the structural evolution observed in nanocrystalline Cu is not material specific but rather is generally observed for face-centered cubic nanocrystalline materials. This means that the observations made here for Cu also shed light on the understanding of wear behavior in other nanocrystalline metals.

Figures 6 and 7 only give a qualitative description of the wear-driven structural evolution as a function of sliding cycle, so we use the GTA analysis code [24] to perform a more quantitative analysis of microstructure. Figure 8 shows the grain structure, with grains colored according to their grain size, after sliding contact under two different applied normal loads. At both loads, the grain size increases with the sliding cycles at the upper part of the sample near the contact surface and the coarsened grains are bounded in many cases by straight twin boundaries. This indicates that the upper part of the sample evolves towards a low-energy state by both reducing the total area of grain boundaries through grain growth and forming low-energy boundaries through deformation twinning. The coarsened region forms a distinct layer on the sample surface and grows into the sample with more sliding cycles.

To show the depth dependence of structural evolution, the average grain size along the Y direction is plotted in Figure 9(a). The inset to this figure shows how the average grain size is calculated. At a given Y position, a cross-sectional plane that is perpendicular to the Y axis is defined, with its projection in XY plane indicated by the red line. The average grain size at a given Y position is then calculated as the averaged diameter of all of the grains that meet this cutting plane. We find that the average grain size does not change from the sample bottom (Y = 0 nm) to a Y position of ~37 nm, even after many sliding cycles, proving that the coarsening is limited to the near surface region. The grain size distribution increases significantly with increasing number of sliding cycles above this region (Y = 37 – 50 nm), although repetitive sliding is required as the first cycle gives no discernable evolution. When a normal load of 481 nN is applied, grain size starts to increase at a Y position of ~42 nm in an approximately linear fashion, followed by a plateau in grain size at a position 5 nm away from the sample surface. A transition region exists that has a gradient in grain size between the coarsened top layer and the matrix with a finer grain size. Increasing the number of sliding cycles moves the curve upward and to the left, demonstrating that repetitive scratching causes the grain size near the sample surface to become larger and pushes the coarsened region deeper into the sample. The slope of the gradient region increases with increasing number of sliding cycles, making the transition region harder to demarcate. This finding suggests that the gradient structure is an intermediate state and can perhaps explain why this structure is not always observed in experiments that characterized the microstructure after many cycles. The comparison between two different normal loads shows that a higher applied load causes more grain growth for a given Y position, consistent with experimental observations [11]. In all of the simulations, regardless of the applied load, the average grain size tends to saturate to a consistent value of ~10 nm near the surface of the sample, although a deeper coarsened region is found at higher applied load. The saturated grain size of 10 nm is in the range of 10-15 nm grain size where Vo et al. [42] observed that grain boundary sliding and rotation stop dominating plasticity in nanocrystalline metals. It should be noted that the whole sample is included in the analysis. As the analysis plane moves near the sample top, fewer grains meet the plane and the averaged grain size gradually loses statistics, leading to a rapid increase in the average grain size after its saturation shown in Figure 9(a).

The sliding-induced structural evolution can also be quantified from another perspective because the fraction of crystal atoms in polycrystalline materials increases with increasing grain size. Figure 9(b) shows the distribution of the number density of face-centered cubic atoms, from the bottom to the top of the sample, including those in wear debris if there is any, after different number of sliding cycles. The number density at a given Y position is measured by counting the number of face-centered cubic atoms inside a 0.2 nm thick bin centered at this position. Similar to Figure 9(a), a coarsened layer on the sample surface, a gradient region, the effect of normal load, and the tendency for grain size saturation with increasing number of sliding cycles is also found. The only major difference is that the number density of crystalline atoms quickly decreases very close to the top surface, a feature that can be used to monitor the location of the surface. Figure 10 shows the number density of crystalline atoms as a function of Y position for a sample worn under an applied normal load of 561 nN (contact pressure of 4.39 GPa right before sliding). Grain growth was also observed on top of the sample, as indicated by the increase in the number density of face-centered cubic atoms. However, the sample surface also moves downward, showing how material is removed by the indenter during the sliding process. This scenario is in sharp contrast to that of diamond and related materials in which wear occurs through surface amorphization followed by material removal [34,43,44].

The grain coarsening and material removal can also be correlated with the local stress distribution within the sample during the indentation and sliding process. Figure 11 shows the von Mises or equivalent stress distribution inside the sample under normal loads of 481 nN and 513 nN. The six components of the stress tensor at a given location are calculated by averaging the corresponding components of all the atoms residing within a cylinder centered at this point that runs through the sample thickness and has a radius of 0.4 nm. The von Mises stress is then calculated based on the averaged stress components at each point. The relatively high stress regions (>1 GPa) are distributed beneath the indenter tip, marking the deformed zone. During the sliding process, the stress moves across the top region, leaving a trail of highly stressed regions related to the plastic deformation that has occurred. However, the most highly stressed region moves with the indenter tip, located beneath and slightly in front of the contact. For both normal loads shown in this figure, the highly stressed region extended approximately 15 nm into the sample, as measured from the original surface location. Alternatively, this would be approximately 35 nm from the bottom of the sample, as marked in Figure 11 by white dotted lines. This depth is close to that of grain growth region shown in Figure 9, again indicating a strong correlation between high-stress and grain growth. The higher normal load results in a wider high stress region, not one that is noticeably deeper. This larger high stress region of influence means a larger region of plastic deformation and grain growth, explaining the observation that grain coarsening occurs more quickly with a 513 nN normal load. It is worth noting that the high-stress region always hovers at the top region during the sliding at the two loads. However, this is no longer the case when increasing the load to 529 nN. As shown in Figure 12, the high stress region becomes larger after the first cycle when compared with the two lower loads and starts to extend much deeper into the sample by the third cycle. The extension of the high stress region and thus plastic deformation, in addition to the removal of the affected region over time, means grain size may never have a chance to reach a distinct, saturated value.

When combined, Figures 9-12 give a complete picture of the effect of applied normal load on structural evolution. When the load is very low, the indenter can only induce elastic deformation to the sample and no prominent structural evolution would be expected. When the load is high enough to induce plastic deformation, grain growth will happen during the sliding process, with higher loads inducing faster growth. As the applied load is further increased, material near the surface of the sample will be removed during the sliding process. Consequently, part of the near-surface region that coarsens is removed by the indenter. Generally, this suggests that any observed structural evolution is a balance of plasticity near the surface and abrasive removal of material.

It should be noted that during the grain growth process, the maximum size of the grown grains becomes similar to the sample thickness. As shown in Figure 13(a), a single grain runs through the thickness direction after 4 cycles under a normal load of 481 nN. Such an observation could suggest that the maximum grain size observed during the scratching process is limited by the sample thickness. To explore this question, another simulation with doubled sample thickness (30 nm) and a doubled normal load (962 nN) was performed. The higher load combines with the larger contact area to give the same contact stress as the prior simulation setup. The side view of the sample after 7 sliding cycles is shown in Figure 13(b), where multiple grains are now observed through the sample thickness. The largest single grain observed near the surface in this figure is ~14 nm in diameter, close to the value (16 nm) obtained in the simulations run with a thinner sample. The number density distribution of face-centered cubic atoms in this sample is plotted together with that of the thin sample in Figure 14. The two curves show a consistent trend, with the only noticeable variation being an increase in noise of the thinner sample, likely due to reduced statistics from the smaller number of atoms measured at a given Y position. Since the structural evolution in both the thin and thick samples develops in a consistent way, the observations of structural evolution do not appear to be related to limited sample thickness but rather show an intrinsic behavior of this nanocrystalline material.

It is expected that the mechanical properties of the sample surface will change as structural evolution occurs, especially those that are closely related to wear properties such as hardness and friction coefficient. To measure hardness, the contact area of the indenter with the sample was first calculated as the area on the cylindrical indenter surface that has neighboring atoms within a distance of 0.1 nm. The hardness at this position was then calculated by dividing the normal load by the projected contact area on the plane perpendicular to the Y axis. The hardness during a sliding cycle is measured by averaging the instantaneous hardness, taken at sliding displacement increments of 0.2 nm. Figure 15(a) shows the surface hardness at two loads as a function of the number of sliding cycles. The hardness generally increases with increasing number of sliding cycles, eventually saturating to a constant value of ~6.5 GPa. This wear-induced surface hardening is consistent with coarsening near the sample surface, as inverse Hall-Petch behavior is expected for Cu with average grain sizes below ~11 nm [45]. During the sliding process, high-energy random grain boundaries on the sample top are gradually replaced by low-energy coherent twin boundaries, which may also contribute to hardening [46]. The correlation between hardness and evolved surface grain size is consistent with the hardness dependence at small grain sizes shown by Li and Szlufarska [19], most likely because grain boundary activities are involved in sliding in both simulations. The friction coefficient was also calculated by dividing the frictional force with the normal load, with measurements again averaged over each sliding cycle. We do not include friction data from the first cycle, because the friction force is zero at the very beginning and gradually increases with sliding distance at the beginning of the sliding process. Figure 15(b) shows that the friction coefficient decreases with increasing number of sliding cycles for both applied normal loads and tends to saturate to a constant value as well, inversely proportional to the hardness trend. Under a constant normal load, an increase in hardness leads to a decrease in contact area between the indenter and the sample surface, which in turn decreases the friction coefficient. It is worth noting that the interaction between the indenter and surface atoms is purely repulsive and thus adhesion-free. The friction force thus comes only from the interaction between the indenter and the atoms ahead, with atoms right below and behind the indenter contribute nothing to the force. A higher normal load pushes deeper into the material and puts more atoms in front of the moving indenter, resulting in higher lateral force in the moving direction. Friction coefficient will increase when the increase of the resultant lateral force outpaces the change to the applied normal force, which explains why the friction coefficient at 513 nN is always higher than that at 481 nN, even though the hardness at the two loads is nearly identical. This wear scenario is similar to the abrasive wear that dominates when a hard surface slides on a softer surface. Despite the lack of adhesion, the correlation of friction coefficient with both surface grain size and applied normal load is consistent with the results shown by Li and Szlufarska [19], who did include adhesion in their simulations. This suggests that abrasive wear dominates the sliding when a large amount of plowed material piles up in front of the indenter.

Both grain coarsening and the formation of low-energy twin boundaries suggest that plasticity is working to lower the energy state of the system. One may also wonder whether grains on top of the sample will preferentially align so that the {111} atomic planes are the surface, since this plane has the lowest surface energy. To test this hypothesis, atomic configurations after 8 sliding cycles at a normal load of 481 nN and 513 nN were plotted in Figure 16, with each grain colored according to the lowest relative angle between all <111> axes of this grain with the global Y axis or surface normal. In other words, a grain with a zero relative angle would have a {111} plane aligned perfectly with the Y axis. Figure 16 shows that, while some surface grains have orientations close to {111}, many other do not have such an orientation. Therefore, we find that surface energy is not a dominant driving force for evolution to a lower energy state, a finding consistent with experimental observations that grains show no texture in evolved layers from wear experiments [9]. This observation is different from what was observed by Romero et al., where a preferential orientation of coarsened grains did manifest near the contact interface [21]. This difference can perhaps be associated with the different simulation setups, with our setup more closely matching experimental conditions and therefore reproducing the lack of texture from experiment.

## IV. CONCLUSIONS

In this work, the wear-driven structural evolution of nanocrystalline metals was studied as a function of sliding cycle and applied load using molecular dynamics simulations and quantitative analysis was performed with the Grain Tracking Algorithm. The normal load of the sliding indenter, instead of its surface displacement, is kept constant to better mimic experimental conditions. Our results show that localized plastic deformation occurs around the indenter, with partial dislocations and deformation twinning dominating plasticity. Grain growth also occurs near the contact surface, first forming a gradient structure and then saturating to a distinct layer of coarsened grains. Higher applied normal loads promote faster grain growth, although the material that has evolved can be removed by abrasive wear if the normal load is too high. Structural evolution at the sample surface leads to an increase in surface hardness and a decrease in friction coefficient. Although the imposed stress by the scratching indenter tends to push the system into the lowest potential energy state with fewer grains and more twin boundaries, we did not observe a preferred texture in the surface grains. As a whole, this work improves the mechanistic understanding of nanocrystalline wear, while also providing a methodology for simulating sliding wear in a way that can more readily be compared to experiments.

## ACKNOWLEDGEMENTS

This research was supported by National Science Foundation under award No. CMMI-1462717. Z.P. acknowledges help from Dr. Jason Panzarino with running the Grain Tracking Algorithm (GTA) code.

# REFERENCES

[1] A. R. Jones, J. A. Hamann, A. C. Lund, and C. A. Schuh, Plating and Surface Finishing **97**, 52 (2010).

[2] J. Schiotz, F. D. Di Tolla, and K. W. Jacobsen, Nature **391**, 561 (1998).

[3] H. Van Swygenhoven and P. M. Derlet, Phys. Rev. B **64**, 224105 (2001).

[4] K. S. Kumar, S. Suresh, M. F. Chisholm, J. A. Horton, and P. Wang, Acta Mater. **51**, 387 (2003).

[5] Z. Shan, E. A. Stach, J. M. K. Wiezorek, J. A. Knapp, D. M. Follstaedt, and S. X. Mao, Science **305**, 654 (2004).

[6] V. Yamakov, D. Wolf, S. R. Phillpot, A. K. Mukherjee, and H. Gleiter, Nat. Mater. **3**, 43 (2004).

[7] A. Emge, S. Karthikeyan, and D. A. Rigney, Wear **267**, 562 (2009).

[8] T. J. Rupert, W. Cai, and C. A. Schuh, Wear **298–299**, 120 (2013).

[9] T. J. Rupert and C. A. Schuh, Acta Mater. **58**, 4137 (2010).

[10] J. F. Archard, Journal of Applied Physics **24**, 981 (1953).

[11] N. Argibay, T. A. Furnish, B. L. Boyce, B. G. Clark, and M. Chandross, Scripta Mater. **123**, 26 (2016).

[12] H. Q. Sun, Y. N. Shi, and M. X. Zhang, Wear **266**, 666 (2009).

[13] H. A. Padilla II, B. L. Boyce, C. C. Battaile, and S. V. Prasad, Wear **297**, 860 (2013).

[14] S. V. Prasad, C. C. Battaile, and P. G. Kotula, Scripta Mater. **64**, 729 (2011).

[15] B. Yao, Z. Han, Y. S. Li, N. R. Tao, and K. Lu, Wear **271**, 1609 (2011).

[16] J. Zhang, T. Sun, Y. Yan, D. Shen, and X. Li, Journal of Applied Physics **112**, 073526 (2012).

[17] J. Li, B. Liu, H. Luo, Q. Fang, Y. Liu, and Y. Liu, Computational Materials Science **118**, 66 (2016).

[18] N. Beckmann, P. A. Romero, D. Linsler, M. Dienwiebel, U. Stolz, M. Moseler, and P. Gumbsch, Physical Review Applied **2**, 064004 (2014).

[19] A. Li and I. Szlufarska, Phys. Rev. B **92**, 075418 (2015).

[20]   M. Mishra, C. Tangpatjaroen, and I. Szlufarska, Journal of the American Ceramic Society **97**, 1194 (2014).

[21]   P. A. Romero, T. T. Järvi, N. Beckmann, M. Mrovec, and M. Moseler, Phys. Rev. Lett. **113**, 036101 (2014).

[22]   T. J. Rupert, Journal of Applied Physics **114**, 033527 (2013).

[23]   J. Panzarino, F., J. Ramos, J., and T. Rupert, J., Modelling and Simulation in Materials Science and Engineering **23**, 025005 (2015).

[24]   J. Panzarino and T. Rupert, JOM **66**, 417 (2014).

[25]   J. F. Panzarino, Z. Pan, and T. J. Rupert, Acta Mater. **120**, 1 (2016).

[26]   S. Plimpton, Journal of Computational Physics **117**, 1 (1995).

[27]   Y. Mishin, M. J. Mehl, D. A. Papaconstantopoulos, A. F. Voter, and J. D. Kress, Phys. Rev. B **63**, 224106 (2001).

[28]   A. Stukowski, Modelling and Simulation in Materials Science and Engineering **18**, 015012 (2010).

[29]   J. D. Honeycutt and H. C. Andersen, J. Phys. Chem. **91**, 4950 (1987).

[30]   A. Stukowski, Modelling and Simulation in Materials Science and Engineering **20**, 045021 (2012).

[31]   C. L. Kelchner, S. J. Plimpton, and J. C. Hamilton, Phys. Rev. B **58**, 11085 (1998).

[32]   J. Knap and M. Ortiz, Phys. Rev. Lett. **90**, 226102 (2003).

[33]   E. T. Lilleodden, J. A. Zimmerman, S. M. Foiles, and W. D. Nix, Journal of the Mechanics and Physics of Solids **51**, 901 (2003).

[34]   X. Hu, M. V. P. Altoe, and A. Martini, Wear **370–371**, 46 (2017).

[35]   Y. Geng, J. Zhang, Y. Yan, B. Yu, L. Geng, and T. Sun, PLOS ONE **10**, 0131886 (2015).

[36]   Z.-D. Sha, V. Sorkin, P. S. Branicio, Q.-X. Pei, Y.-W. Zhang, and D. J. Srolovitz, Appl. Phys. Lett **103**, 073118 (2013).

[37]   J. Y. Wu, S. Nagao, J. Y. He, and Z. L. Zhang, Nano Letters **11**, 5264 (2011).

[38]   A. Cao and Y. Wei, Phys. Rev. B **74**, 214108 (2006).

[39]   X. Z. Liao, Y. H. Zhao, S. G. Srinivasan, Y. T. Zhu, R. Z. Valiev, and D. V. Gunderov, Appl. Phys. Lett **84**, 592 (2004).

[40]     Y. T. Zhu, X. Z. Liao, and R. Z. Valiev, Appl. Phys. Lett **86**, 103112 (2005).

[41]     Y. Mishin, D. Farkas, M. J. Mehl, and D. A. Papaconstantopoulos, Phys. Rev. B **59**, 3393 (1999).

[42]     N. Q. Vo, R. S. Averback, P. Bellon, S. Odunuga, and A. Caro, Phys. Rev. B **77**, 134108 (2008).

[43]     L. Pastewka, S. Moser, P. Gumbsch, and M. Moseler, Nat Mater **10**, 34 (2011).

[44]     F. M. van Bouwelen, A. L. Bleloch, J. E. Field, and L. M. Brown, Diamond and Related Materials **5**, 654 (1996).

[45]     J. Schiøtz and K. W. Jacobsen, Science **301**, 1357 (2003).

[46]     K. Lu, L. Lu, and S. Suresh, Science **324**, 349 (2009).22

# FIGURES AND CAPTIONS

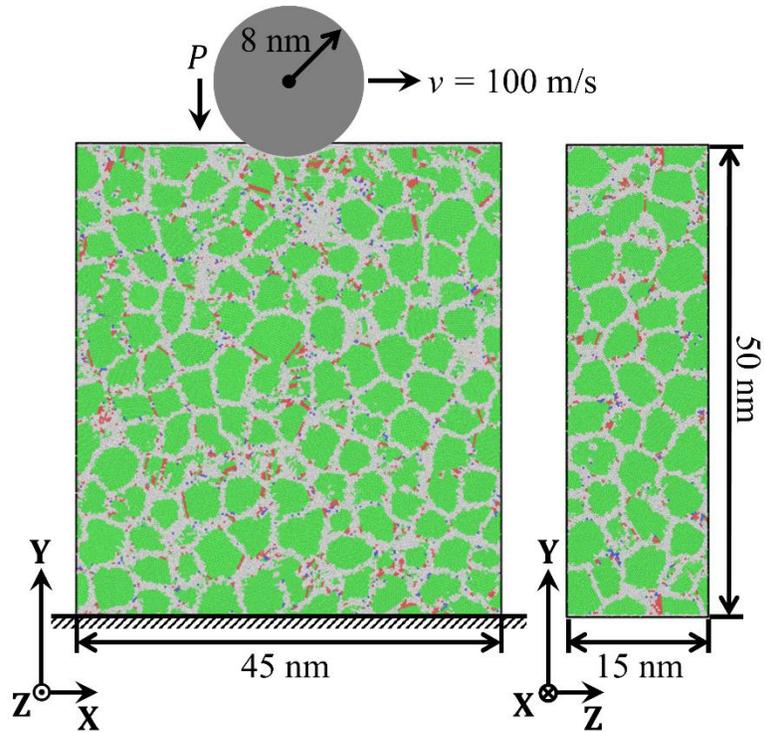

Figure 1. The simulation setup for the indentation and scratching of a nanocrystalline sample with an average grain size of ~5 nm constructed with Voronoi tessellation method. Atoms are colored according to CNA, to show the grain structure.

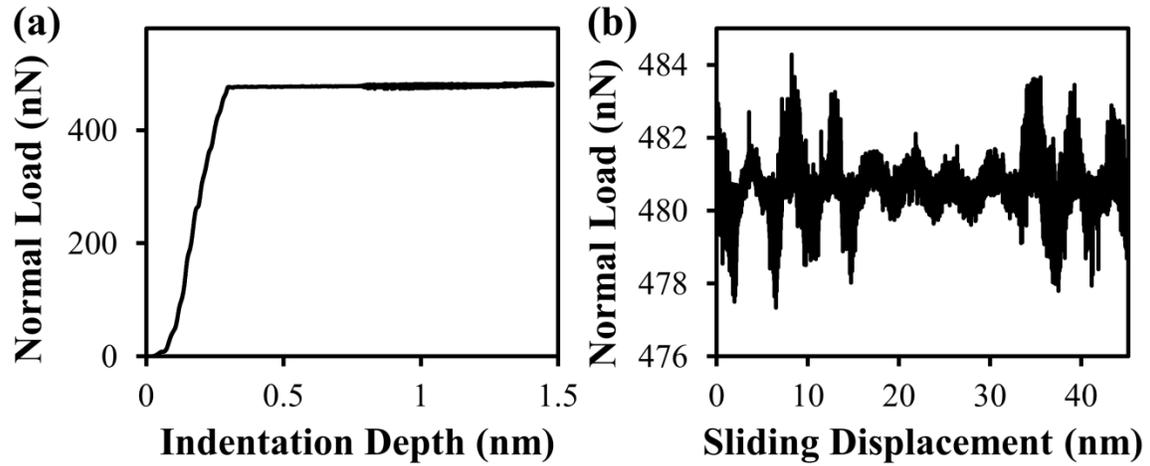

Figure 2. The applied normal load on the indenter as a function of (a) indentation depth during the initial indentation process and (b) sliding displacement during the scratching process. Fluctuations in the applied load can be controlled to within ±1% of the target load during the scratching process.

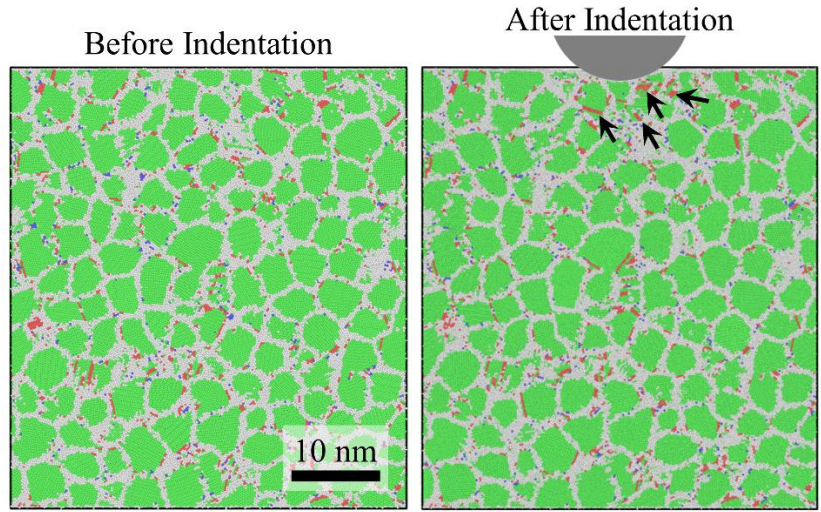

Figure 3. Atomic snapshots taken right before and after indentation with an applied normal load of 481 nN. Atoms are colored according to CNA. The lower part of the indenter is drawn to mark its position after indentation. The black arrows mark the position of newly formed stacking faults due to the indentation.

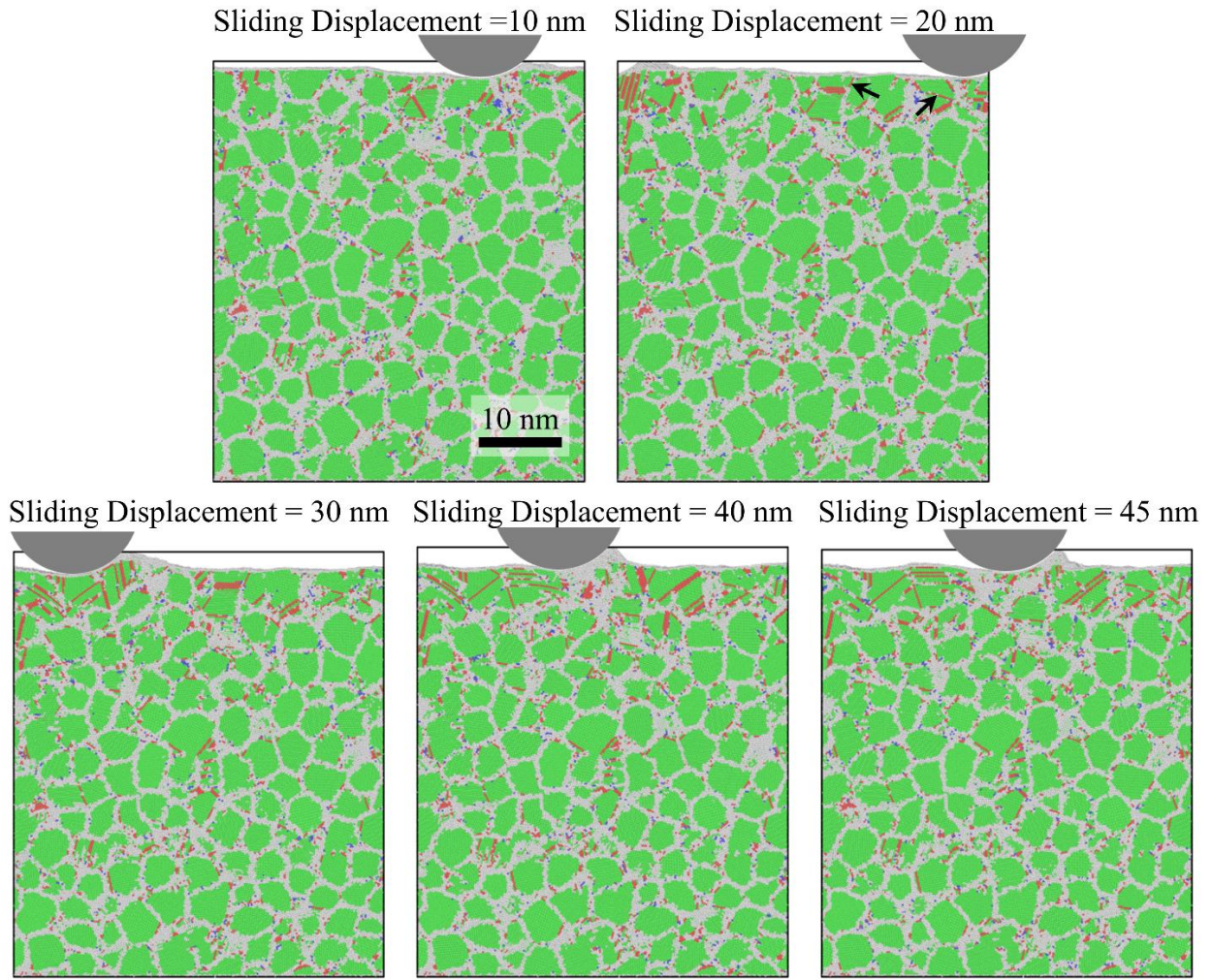

Figure 4. Atomic snapshots taken at different sliding displacements during the first cycle of the scratching process with an applied normal load of 481 nN. Partial dislocations and deformation twinning are observed near the contact surface. The black arrows mark the position of two newly formed twin boundaries.

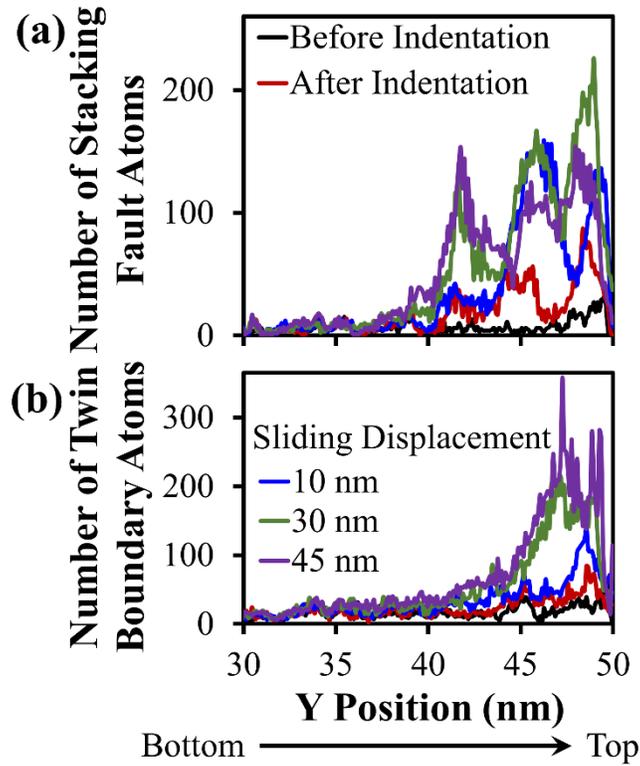

Figure 5. Distribution of (a) stacking fault and (b) twin boundary atoms along the Y position before and after indentation and at different sliding displacements of the first wear cycle in a sample loaded under a normal load of 481 nN. The population of both types of atoms increases with increasing sliding displacement near the contact surface.

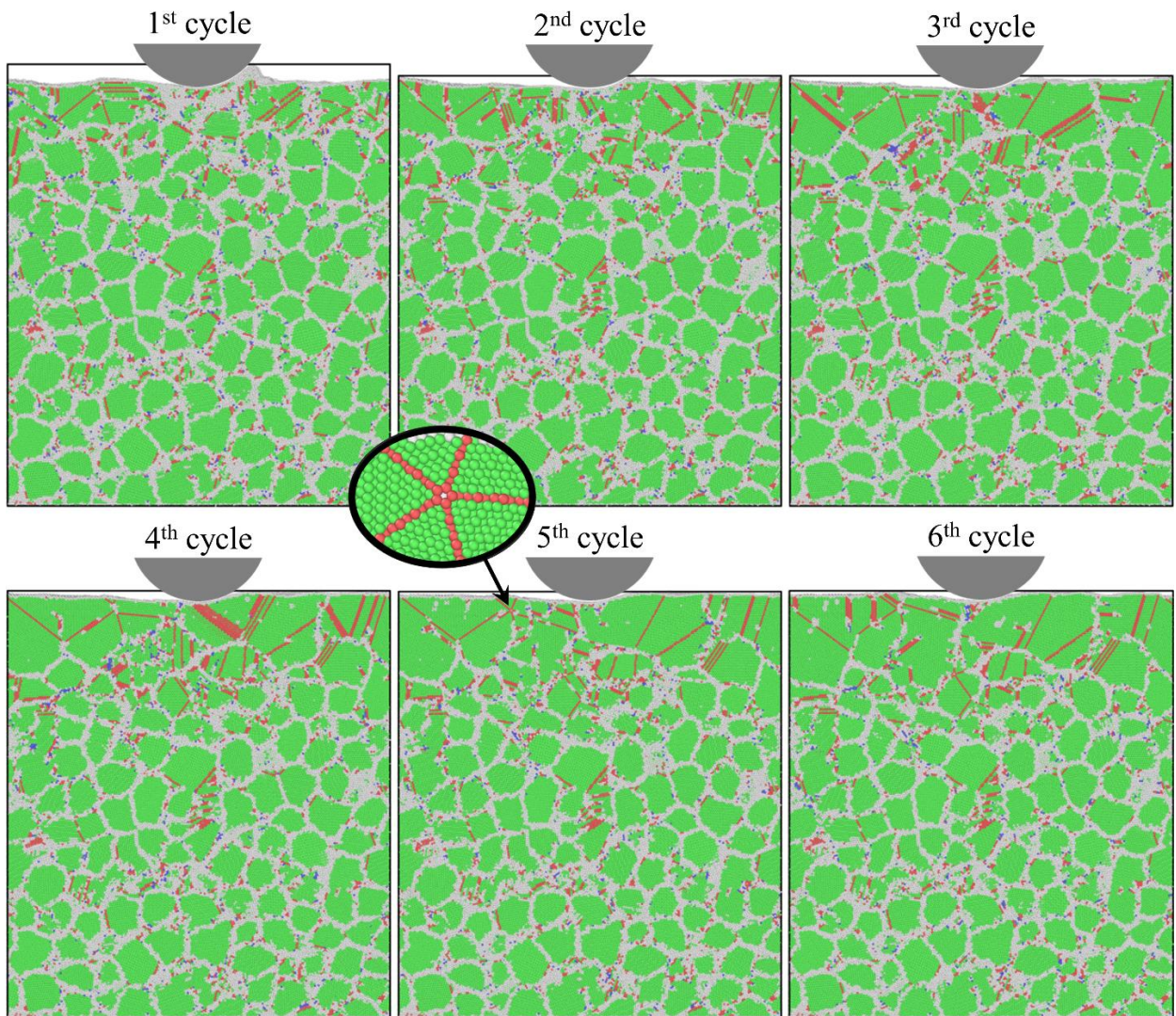

Figure 6. Atomic snapshots after different numbers of sliding cycles under a normal load of 481 nN, showing grain growth and the formation of twin boundaries. An example of a five-fold twin is shown in the inset to this figure.

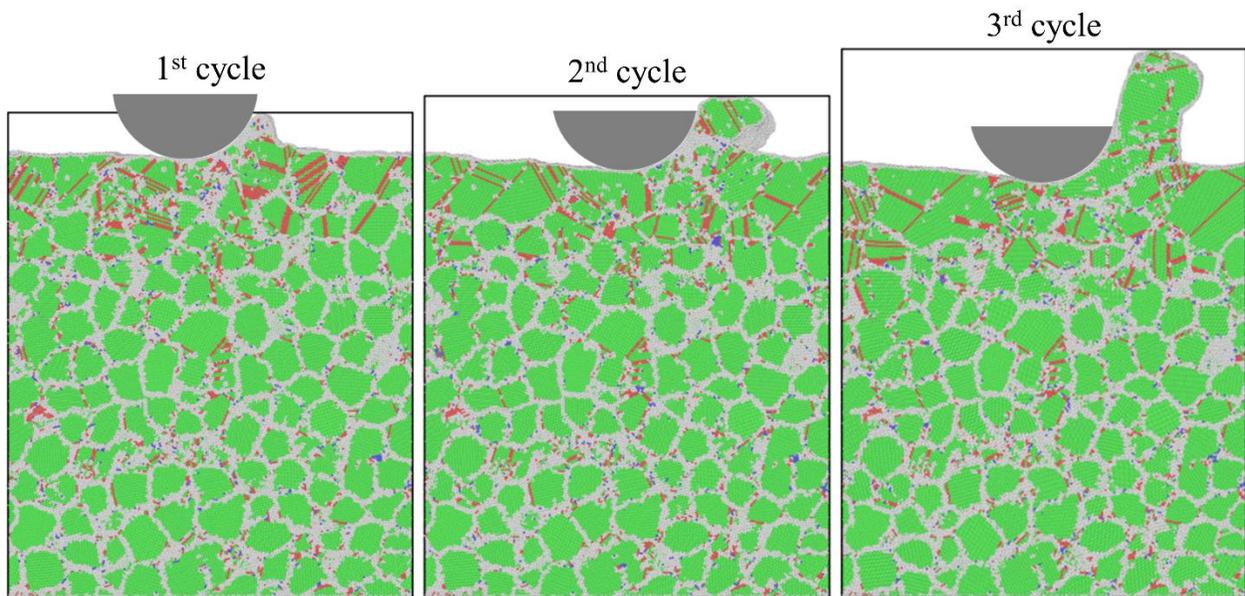

Figure 7. Atomic snapshots after 1–3 sliding cycles under a normal load of 529 nN, showing material removal through an abrasive wear process.

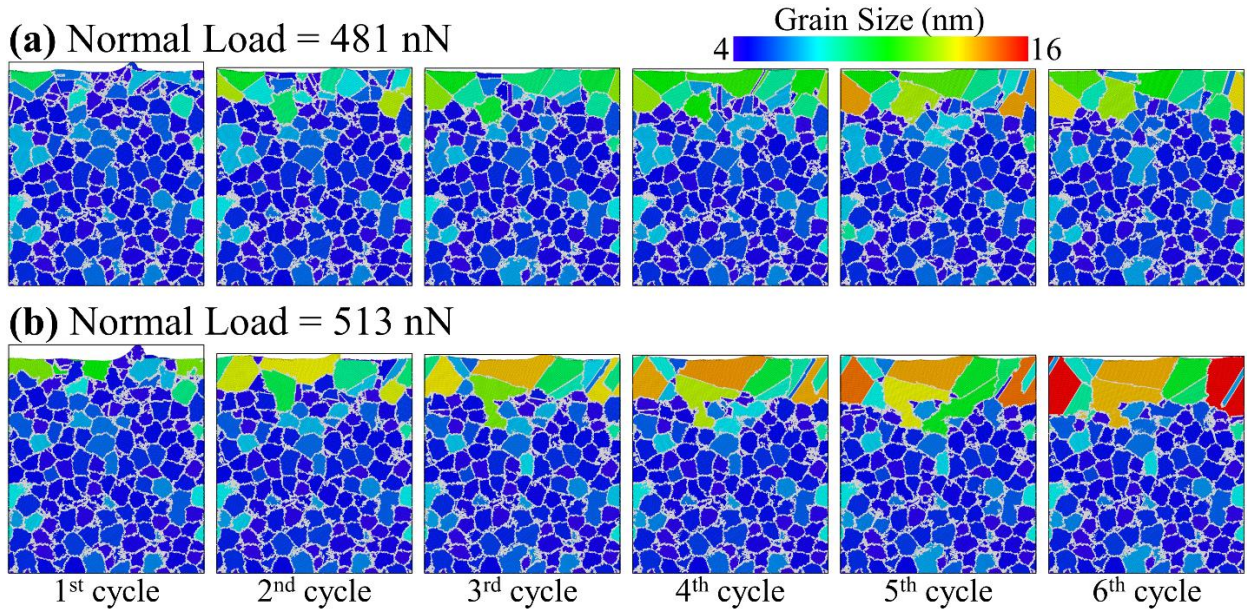

Figure 8. Snapshots after 1–6 sliding cycles under a normal load of (a) 481 nN and (b) 513 nN. Each grain is identified with the Grain Tracking Algorithm (GTA) and colored according to its spherically equivalent diameter. Grain boundary atoms are shown in white.

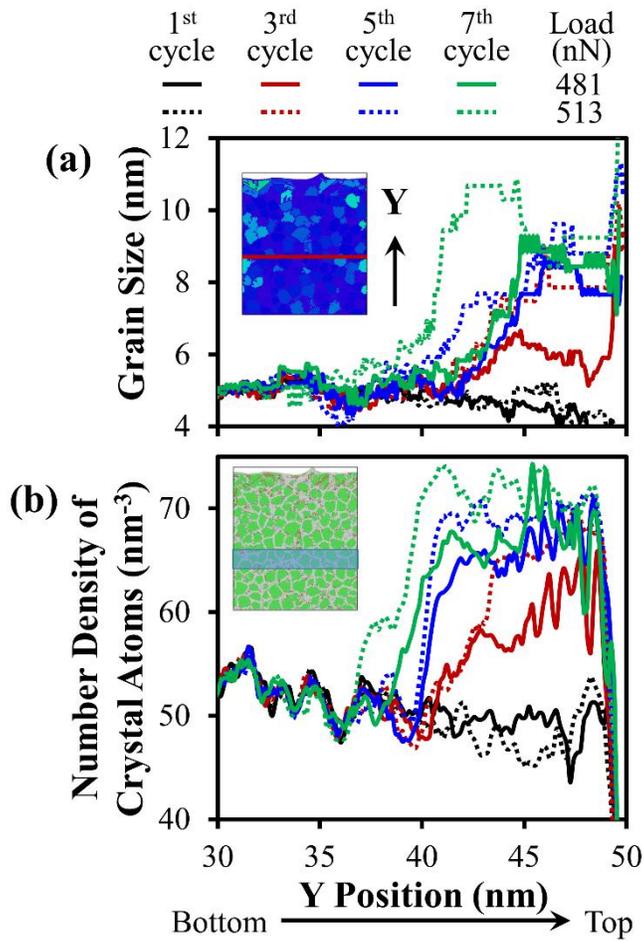

Figure 9. (a) Grain size distribution and (b) number density distribution of face-centered cubic atoms along the Y position in samples after different numbers of sliding cycles under normal loads of 481 nN and 513 nN. Grain growth occurs near the sample surface, while no evolution occurs below ~37 nm. Insets show how the grain size and number density distributions are calculated.

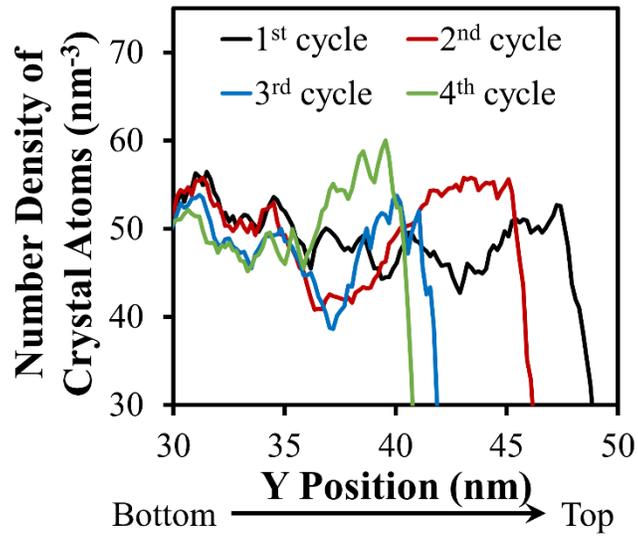

Figure 10. Number density distribution of face-centered cubic atoms as a function of Y position after 1–4 cycles under a normal load of 529 nN. The position of sample surface moves downward with increasing number of sliding cycles, indicating an abrasive material removal process.

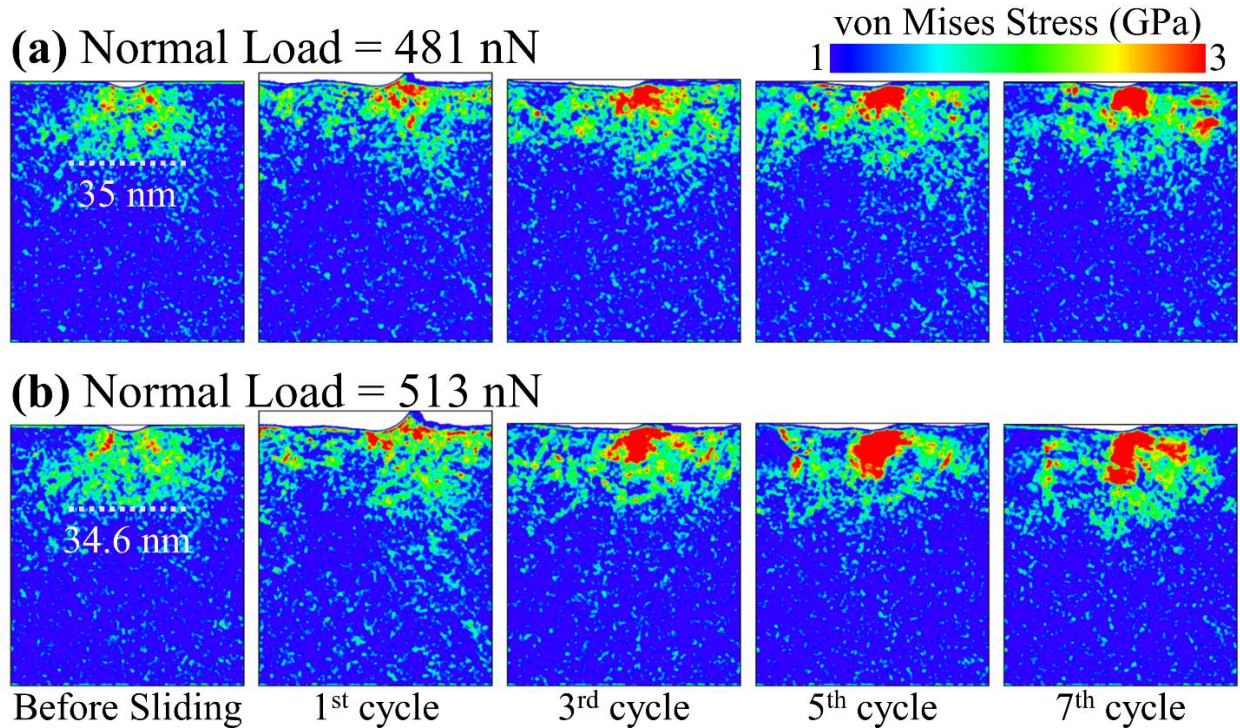

Figure 11. Von Mises stress distribution in samples right before sliding and after different number of sliding cycles under normal loads of (a) 481 nN and (b) 513 nN. The white dotted lines roughly mark the lower position of high-stress region right before sliding begins.

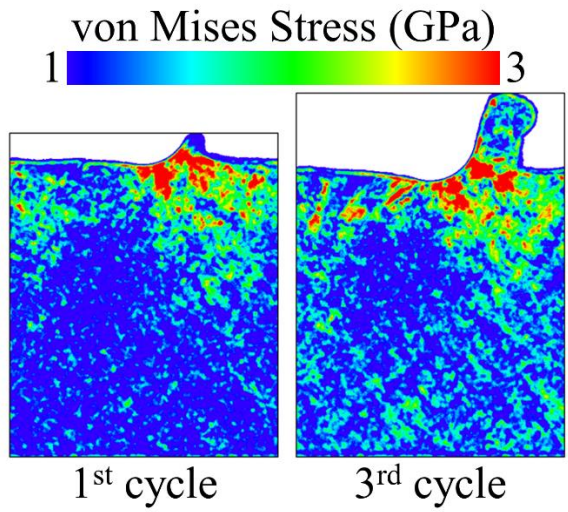

Figure 12. Von Mises stress distribution in a sample after the 1$^{st}$ and 3$^{rd}$ sliding cycles under a normal load of 529 nN.

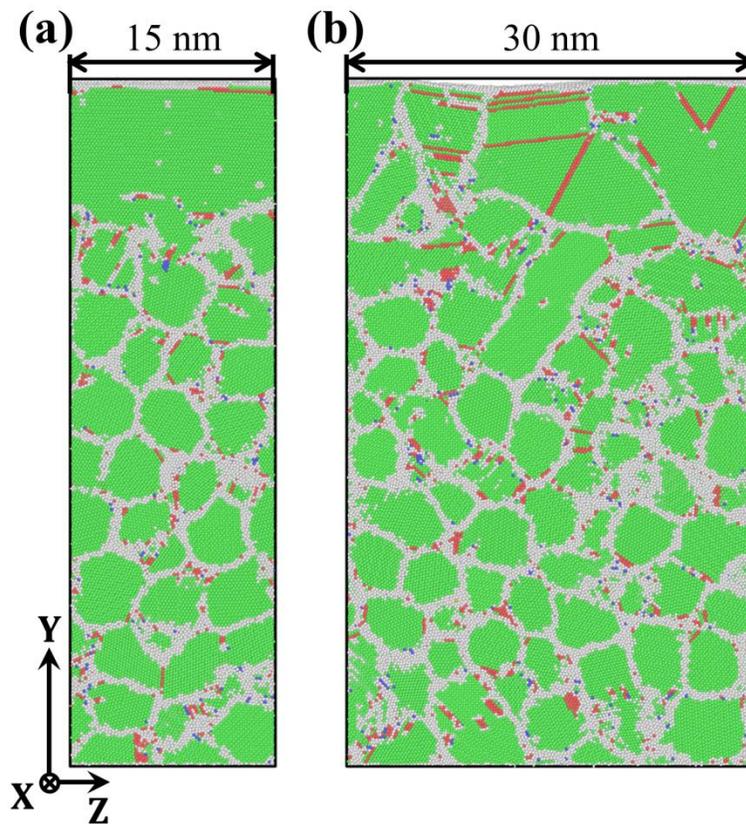

Figure 13. Side view of (a) the 15 nm thick sample after 4 sliding cycles under a normal load of 481 nN and (b) the 30 nm thick sample after 7 sliding cycles under a normal load of 962 nN. Grown grains run through thickness direction in the thin sample but not in the thick sample.

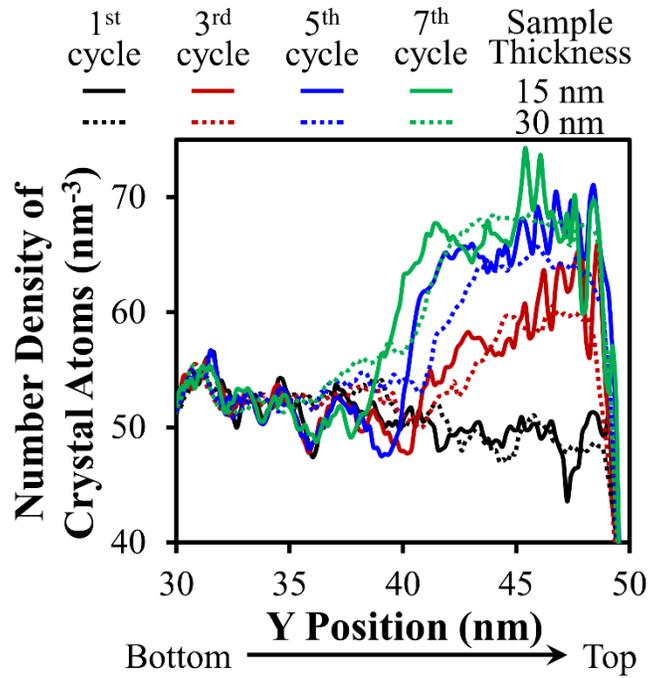

Figure 14. Number density distribution of face-centered cubic atoms along the Y position for the 15 nm thick and 30 nm thick samples. The two samples demonstrate similar evolution despite the very different thickness.

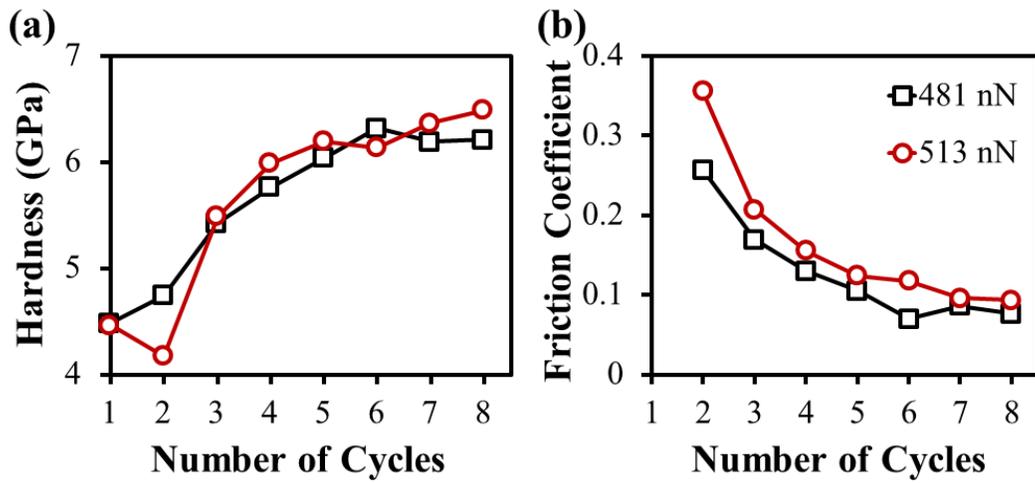

Figure 15. Averaged (a) surface hardness and (b) friction coefficient as a function of the number of sliding cycles, under two applied normal loads. Both properties tend to saturate to constant values after ~6 cycles.

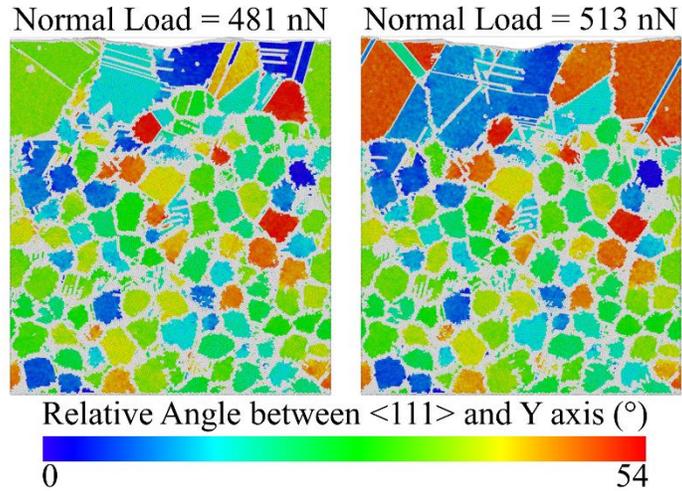

Figure 16. Sample snapshots after 8 wear cycles under normal loads of 481 nN and 513 nN. Each grain is identified with the GTA and colored according to the smallest relative angle between a <111> grain axis and the global Y axis. No preferred texture is found near the surface.